\documentclass{appolb}
\usepackage{epsfig}

\begin{document}
\title{Cluster radioactivity in $^{114}$Ba in the HFB theory%
\thanks{Presented at Zakopane  Conference on Nuclear Physics}%
}
\author{M. Warda
\address{Katedra Fizyki Teoretycznej, Uniwersytet Marii Curie--Sk\l odowskiej,\\
        ul. Radziszewskiego 10, 20-031 Lublin, Poland}
}
\maketitle
\begin{abstract}
Cluster radioactivity in  $^{114}$Ba is described as a spontaneous fission with a large mass asymmetry within the self-consistent HFB theory.  A new fission valley with large octupole deformation is found in the potential energy surface. The fragment mass asymmetry of this fission mode corresponds to the expected one in cluster radioactivity with the emission of  $^{16}$O  predicted with a very long half-life.
\end{abstract}

\vspace{-5mm}
\section{Introduction}
Cluster radioactivity (CR) discovered by Rose and Jones \cite{ros84} was observed in 19 actinides. This is an exotic process in which a light nucleus ($^{14}$C up to  $^{34}$Si) is emitted and a nucleus from the region of  $^{208}$Pb constitutes the heavy mass residue. Since a doubly-magic structure of the heavy fragment is a necessary condition for CR, such a decay may be also associated with the doubly-magic $^{100}$Sn. 
The shortest CR half-lives in actinides  are obtained for the emission of a light $^{14}$C cluster. This implies that the best candidate as emitter from the $^{100}$Sn region is $^{114}$Ba. Experimental investigations of cluster decay in this isotope have been carried out, but no evidence  was found for  CR and only a lower limit for the  half-life $t_{1/2}$ could be determined. Oganessian et al. \cite{oga94} found $t_{1/2}>10^3$ s, whereas the value found by  Guglielmetti et al. \cite{gug95,gug96} is larger $t_{1/2}>1.2\cdot10^4$ s.

We showed in a fully microscopic approach \cite{rob08a,rob08b} that CR can be described as a very asymmetric spontaneous fission process. The  potential energy surfaces were determined as  function of quadrupole and octupole moments within the framework of the Hartree-Fock-Bogolubov model with the D1S Gogny force \cite{war02}.  New fission valleys in the energy surfaces with fragment mass asymmetry characteristic for CR were found.  The experimental half-lives of CR in actinides were reproduced with good accuracy. The same model can be applied to lighter systems. In this contribution we present the results of our investigations of CR in $^{114}$Ba. 

\vspace{-5mm}
\section{Results}

\begin{figure}
\begin{center}
\includegraphics[angle=270,width=0.8\columnwidth]{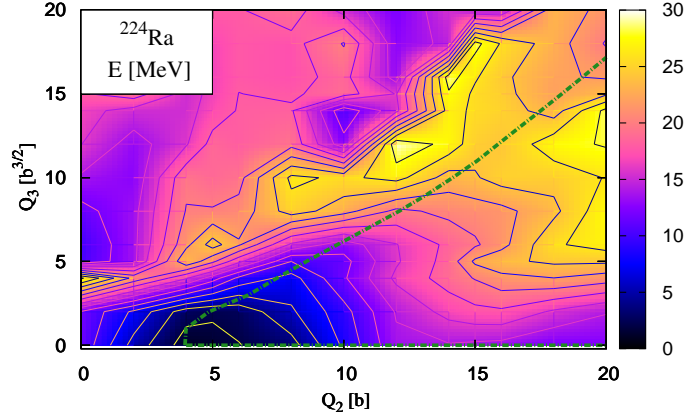}
\caption{\label{barq2} Potential energy surface in $^{114}$Ba as function of quadrupole moment $Q_2$ and octupole moment $Q_3$. Lines of constant energy are plotted every 2 MeV. Bold dash-dotted lines are plotted along the two different fission paths.}
\end{center}
\end{figure}

\begin{figure}
\begin{center}
\includegraphics[angle=0,width=0.6\columnwidth]{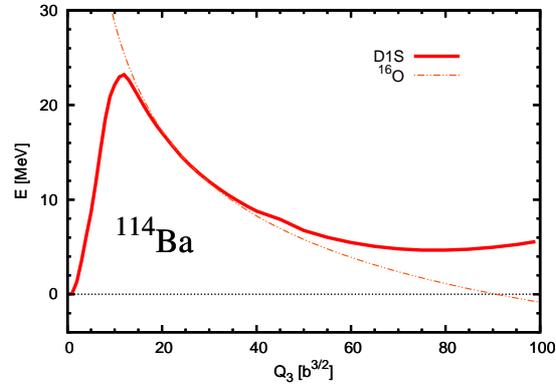}
\caption{\label{barq2} Asymmetric fission barrier in $^{114}$Ba  as function of the  octupole moment $Q_3$. The thin dashed line corresponds to the classical Coulomb energy of two fragments $^{16}$O and $^{98}$Cd.}
\end{center}
\end{figure}
\begin{figure}
\begin{center}
\includegraphics[angle=270,width=0.8\columnwidth]{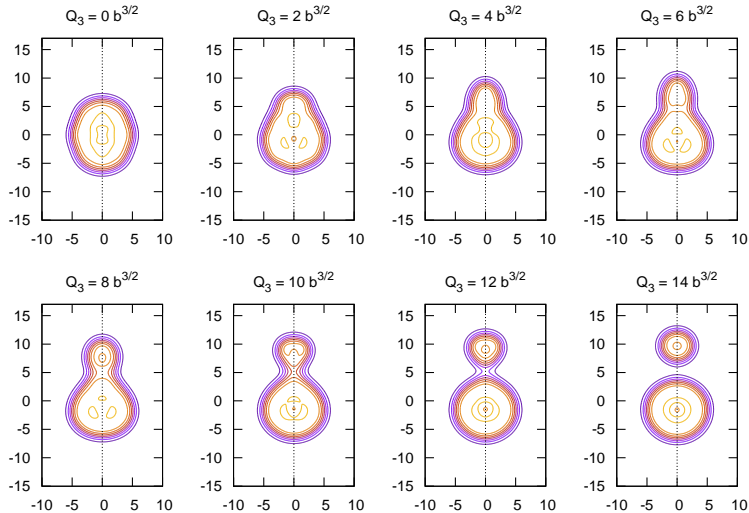}
\caption{\label{barq2} Evolution of the shape of  $^{114}$Ba along the asymmetric fission path. Equidensity lines are plotted every 0.02 fm$^{-3}$.}
\end{center}
\end{figure}

In Fig. 1. the potential energy surface of $^{114}$Ba is shown as function of quadrupole and octupole moments. There are two fission paths on that surface, marked with dashed lines. The first one, along $Q_3=0$, leads to mass symmetric fission. The corresponding barrier is however too large to make spontaneous fission possible. The second fission path starts in the ground state at $(Q_2=4$ b, $Q_3=0$ b$^{3/2})$ and goes through the saddle point at $(Q_2=16$ b, $Q_3=12$ b$^{3/2})$. The nucleus takes shapes with octupole deformation along this path and reproduces mass asymmetry of CR. The asymmetric potential energy barrier is very high and reaches 23 MeV, which corresponds to a typical value of a CR barrier in the actinide region. The neck  rupture takes place at the  saddle point which means that the scission point is localized on top of the barrier. While the Coulomb interaction between fragments decreases with the distance between fragments the energy decreases as well. This can be clearly seen in Fig. 2 where the potential energy along the cluster emission path is plotted as function of the octupole moment. Due to the insufficient size of the basis used in the calculations  at huge deformations, the potential energy of the nucleus at large $Q_3$ is not properly  minimized. Therefore the classical formula for the Coulomb energy is used to determine the fission barrier beyond the  scission point.  The Coulomb energy, marked by a dashed line in Fig. 2, perfectly coincides with the self-consistent energy close to the saddle point and deviates from it significantly for  $Q_3>70$ b$^{3/2}$.

In Fig. 3 contour-plots of the nuclear density distribution along the CR fission path are shown. The molecular structure of the deformed nucleus can be seen  at very low octupole deformation $Q_3>4$ b$^{3/2}$. The neck becomes thinner and nuclear density in the neck  decreases gradually with increasing deformation up to the saddle point at $Q_3=12$ b$^{3/2}$. Finally at $Q_3=14$ b$^{3/2}$ the neck is ruptured and two fragments are observed. A further increase of the deformation of the system is obtained with increasing distance between the fragments. 
The light cluster $^{16}$O and the  heavy mass residue  $^{98}$Cd are created at the scission point.  The discrepancy of two protons  from expected $^{14}$C and  $^{100}$Sn fragments is noticed.  Due to strong shell effects the doubly-magic structure of the cluster is energetically more favourable than the doubly-magic heavy fragment.  Similarly nuclei in the vicinity of $^{208}$Pb are also often observed as one of the products of the CR decay of actinides.

The half-life of CR in $^{114}$Ba has been calculated in the  WKB approximation \cite{war02}. The octupole moment has been taken as leading coordinate in an action integral \cite{rob08a,rob08b}.  Microscopic values of zero-point energy along the whole fission path have been used to calculate the action integral.  A half-life of  $t_{1/2}= 4.7 \cdot 10 ^9$ s has been obtained. This value fits experimental data, being much larger than the experimental lower limit. 

\vspace{-5mm}
\section{Conclusions}
A new valley in the potential energy surface of $^{114}$Ba leading to fission with two fragments  $^{16}$O and $^{98}$Cd is discovered. The decay products correspond to the cluster radioactivity  predicted in this nucleus. The fission barrier for this decay is 23 MeV  high.  The nucleus has octupole deformation along the whole fission path. A large half-life for cluster radioactivity $t_{1/2}= 4.7 \cdot 10^9$ s is found which is consistent with experimental data. 

This work is partially supported by Grant No.\ N~N202~231137 from MNiSW (Poland). The author is grateful to L.M. Robledo for helpful discussion at all stages of this work.


\end{document}